\documentclass[aps,pre,reprint,amsmath,amssymb,superscriptaddress,showpacs,floatfix]{revtex4-2}

\usepackage{graphicx}
\usepackage{dcolumn}
\usepackage{bm}
\usepackage[colorlinks, allcolors=blue]{hyperref}
\usepackage[usenames, dvipsnames]{color}
\usepackage{float}

\usepackage{lipsum}
\usepackage{titlesec}
\titlespacing\section{0pt}{12pt plus 4pt minus 4pt}{1pt plus 20pt minus 2pt}
\usepackage{xcolor}
\usepackage{tabularx}
\usepackage{amsmath}
\usepackage{comment}
\usepackage{afterpage}
\usepackage{placeins}
\usepackage{url}
\usepackage{booktabs}
\usepackage{multirow}
\usepackage{array}
\usepackage{setspace}
\graphicspath{{Figs/}}
\usepackage{siunitx}
\usepackage{hhline}
\usepackage{xfrac}
\usepackage{float,graphicx}
\usepackage{mathtools}
\usepackage{listings}
\usepackage{amssymb}
\usepackage[normalem]{ulem}
\usepackage{titlesec}
\usepackage{hyperref}
\usepackage{amsfonts}
\usepackage[version=4]{mhchem}

\usepackage{epstopdf}
\catcode`@11
\def\seceqaa{\@addtoreset{equation}{section}
principles\def\theequation{A\arabic{equation}}}
\def\seceqbb{\@addtoreset{equation}{section}
\def\theequation{B\arabic{equation}}}
\def\seceqcc{\@addtoreset{equation}{section}
\def\theequation{C\arabic{equation}}}
\def\seceqdd{\@addtoreset{equation}{section}
\def\theequation{D\arabic{equation}}}
\def\seceqee{\@addtoreset{equation}{section}
\def\theequation{E\arabic{equation}}}
\def\seceqff{\@addtoreset{equation}{section}
\def\theequation{F\arabic{equation}}}
\def\seceqgg{\@addtoreset{equation}{section}
\def\theequation{G\arabic{equation}}}
\def\seceqhh{\@addtoreset{equation}{section}
\def\theequation{H\arabic{equation}}}
\catcode`@11

\DeclareUnicodeCharacter{2009}{\,}

\newcommand{\SINP}{\affiliation{Condensed Matter and Surface Physics Division, Saha Institute of Nuclear Physics, 1/AF 
Bidhannagar, Kolkata 700064, India}}

\newcommand{\BATH}
{\affiliation{Department of Physics, University of Bath, United Kingdom}}

\newcommand{\DFT}{\affiliation{Department of Physics, National Institute of Technology Silchar, Assam 788010, India}}
  
\newcommand{\IACS}{\affiliation{School of Physical Sciences, Indian Association for the Cultivation of Science, Jadavpur, Kolkata 700032, India}}

\newcommand{\SNCBS}{\affiliation{Department of Condensed Matter and Material Physics, S. N. Bose
National Centre for Basic Sciences, Kolkata 700106, India.}}

\newcommand{\HBNI}{\affiliation{Homi Bhabha National Institute, Training School Complex, Anushakti Nagar, Mumbai, 400094, India}}

\newcommand{\BOSE}{\affiliation{Department of Physical Sciences, Bose Institute, 93/1,
Acharya Prafulla Chandra Road, Kolkata 700009, India.}}
\begin{document}

\title{
Spin-phonon interaction in a symmetry-enforced spin-polarized state} 
\author{Suman Kalyan Pradhan}
\email{sumankalyan.pradhan1@saha.ac.in}
\SINP
\author{Dayal Das}
\altaffiliation{These authors contributed equally to this work}
\SNCBS
\author{Shubham Patel}
\altaffiliation{These authors contributed equally to this work}
\BATH
\author{Subhajit Mahapatra}
\BOSE
\author{Sachin Majee}
\SINP
\HBNI
\author{Dibyendu Majee}
\SINP
\HBNI
\author{Arnab Bera}
\IACS

\author{Achintya Singha}
\BOSE

\author{Snehasish Nandy}
\email{snehasish@phy.nits.ac.in}
\DFT
\author{Samik DuttaGupta}
\email{duttagupta.samik@saha.ac.in}
\SINP
\HBNI
\author{Atindra Nath Pal}
\email{atin@bose.res.in}
\SNCBS

\begin{abstract}
Symmetry-governed magnetic materials have emerged as a promising platform for spintronic functionalities without net magnetization or stray magnetic fields, motivating the exploration of how lattice dynamics couple to symmetry-derived spin-polarized electronic states. Understanding spin-phonon coupling in these systems is therefore essential for uncovering the microscopic origin of spin-lattice interactions and for enabling their control in quantum materials. However, this mechanism remains poorly understood because spin polarization originates from crystal symmetry rather than conventional magnetic order. Here, we address this issue in the $g$-type altermagnet CoNb$_4$Se$_8$ using temperature- and polarization-resolved Raman spectroscopy, complemented by measurements on a structurally analogous Co-deficient compound lacking well-defined long-range magnetic order. We observe pronounced symmetry-selective phonon renormalization across the magnetic transition in CoNb$_4$Se$_8$, while related phonon anomalies persist in the Co-deficient system, demonstrating that the lattice response cannot be explained solely by conventional exchange-striction associated with coherent magnetic ordering. First-principles calculations reveal that spin-orbit coupling establishes a symmetry-dependent interaction channel between lattice vibrations and symmetry-governed electronic states. Our results identify an alternative mechanism for spin-phonon coupling in symmetry-governed magnetic materials and demonstrate that phonons provide a sensitive probe of symmetry-driven spin polarization even without robust magnetic order. More broadly, this work provides a framework for understanding and engineering spin-lattice functionality in symmetry-driven quantum materials, offering design principles for coupling lattice dynamics to spin-polarized electronic states.
\end{abstract}

\maketitle

{\section{Introduction}}

The coupling between magnetic order and lattice excitations in condensed matter, commonly referred to as spin-phonon coupling (SPC) \cite{Cao2008,Vergara,Go}, plays a pivotal role towards emergent phenomena \cite{Calero,Tian2016,Bansal2020,Bozhko2020,Bera2025,Badola,pal2025}, leading to the possible development of future quantum spintronic functionalities and devices. 
In magnetic materials, including ferromagnets (FMs) and antiferromagnets (AFMs), which are classified by the presence or absence of a net macroscopic magnetization \cite{Neel1951,Neel1971}, SPC is most commonly understood within an exchange-striction framework, wherein lattice distortions modulate interatomic exchange interactions between localized spins in real space \cite{Granado,Wu2022,Weber2022}. In this scenario, the SPC strength scales with the spin-spin correlation $\langle \mathbf{S}_i \cdot \mathbf{S}_j \rangle$ \cite{Sun,Lee2010}. On the other hand, the existence of relativistic or orbital effects (viz. Dzyaloshinskii-Moriya \cite{Son2019} and Kugel-Khomskii interactions \cite{Roy,Huang}, etc.) has been shown to significantly influence SPC, suggesting the insufficiency of the purely exchange-driven origin. This raises an important question concerning the role of underlying relativistic, quantum interactions and crystal symmetry governing SPC and its associated optical signatures in symmetry-governed material systems. 

Altermagnets (AMs) \cite{Mazin2021}, which exhibit nonrelativistic spin splitting in momentum space despite vanishing net magnetization \cite{PhysRevX.12.031042,PhysRevX.12.040501,Krempasky2024,Zhou,Cheong2025,Jungwirth2026}, have recently emerged as promising platforms for low-dissipation and stray-field-free spin transport \cite{Bai,Chi,Leraand,Fender2025}. Beyond their spintronic relevance, the symmetry-governed electronic structure of AMs provides a unique opportunity to investigate spin-phonon coupling beyond conventional exchange-driven magnetic systems. 
In this regard, layered and exfoliable altermagnets are particularly attractive because reduced dimensionality can enhance both electronic correlations \cite{Ghosh2021} and symmetry-related interactions. Among the recently identified AMs \cite{Lee2024,Osumi,Gonzalez,Reimers2024,Zeng2024}, CoNb$_4$Se$_8$, a $g$-type altermagnet \cite{Regmi,Sakhya,Candelora,Dale,Devita}, is especially well suited for such investigations, as tuning the Co concentration systematically modifies both the magnetic ground state and crystal symmetry while preserving the underlying hexagonal structural framework \cite{Mandujano}.
\begin{figure*}
\centering
\includegraphics[width=2\columnwidth]{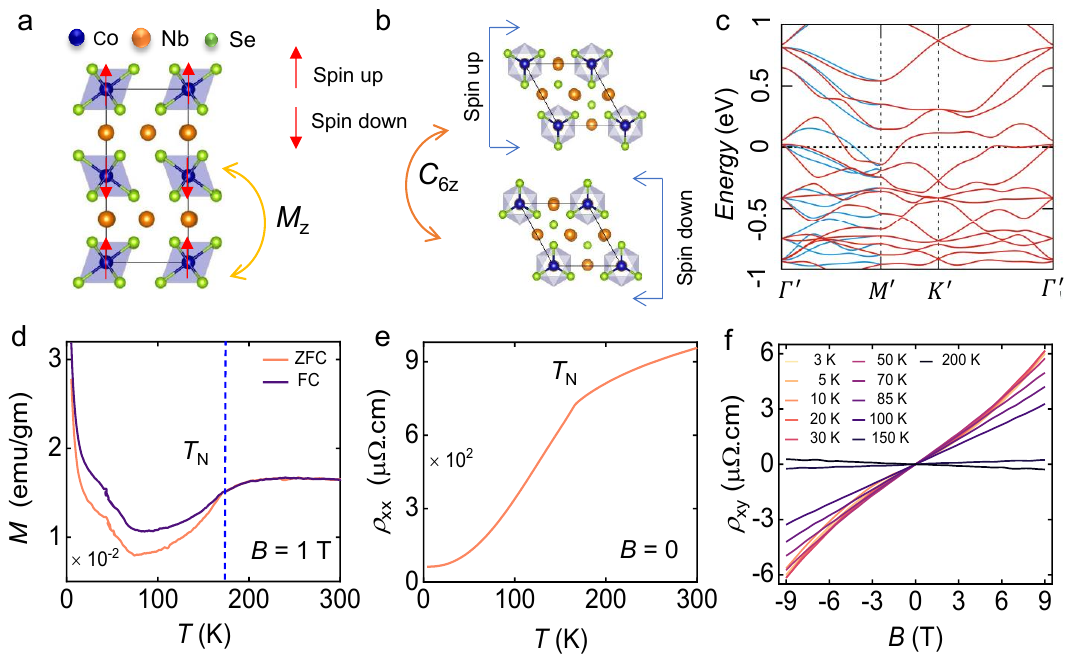}
\caption{
 Magnetic structure of CoNb$_4$Se$_8$ viewed along the (a) $b$ and (b) $c$-axis \cite{Dale}. Red arrows in (a) denote the spin orientations (spin up and spin down) of the magnetic moments. (c) Spin-resolved band structure in the A-type AFM configuration calculated along the high-symmetry path $\Gamma' - M' - K' - \Gamma'$, where the prime denotes the half Brillouin-zone direction along $k_z$. The red and blue curves represent the spin-up and spin-down channels, respectively, and the horizontal dashed line marks the Fermi level. 
(d) Zero-field-cooled and field-cooled magnetization $M(T)$ at $B = 0.1$~T along $c$ axis, showing the onset of long-range order at the N\'eel temperature $T_\text{N}$.
(e) Zero-field longitudinal resistivity ($\rho_\text{xx}$) as a function of temperature. (f) Field-dependent Hall resistivity $\rho_\text{xy}(B)$ at various temperatures.}
\vspace{-0.45cm}
\label{charect}
\end{figure*}
The observation and microscopic understanding of symmetry-governed spin-lattice interactions lie at the heart of the present work. Through temperature- and polarization-resolved Raman spectroscopy, we reveal pronounced symmetry-selective phonon renormalization across the magnetic transition in CoNb$_4$Se$_8$. We additionally find that related phonon anomalies persist in a structurally analogous Co-deficient compound lacking well-defined long-range magnetic order, indicating that the lattice response cannot be fully understood within a conventional exchange-driven framework associated with coherent magnetic ordering. Supported by first-principles calculations, our results further suggest that spin-orbit coupling (SOC) provides a symmetry-dependent pathway linking lattice vibrations to the spin-polarized electronic environment. Together, these findings demonstrate that symmetry-governed electronic interactions can drive an unconventional spin-lattice response that remains robust even when long-range magnetic order is weakened.

{\section{Results and Discussion}}
\subsection{Altermagnetic order}
Single crystals of CoNb$_4$Se$_8$ were synthesized via chemical vapor transport (see Methods), yielding millimeter-sized shiny black platelets. Energy-dispersive X-ray spectroscopy (EDX) measurements performed on multiple crystals confirm the expected chemical composition [Fig.~S1(a), Table~S1]. The high crystalline quality of the as-grown crystals is further evidenced by the sharp diffraction features observed in the two-dimensional X-ray diffraction pattern [Fig.~S1(b)]. Rietveld refinement of powder X-ray diffraction data collected from finely crushed single crystals confirms that CoNb$_4$Se$_8$ crystallizes in a centrosymmetric hexagonal structure with space group $P6_3/mmc$ (No.~194) \cite{Regmi} [Fig.~S1(c), Table~S2]. 
As illustrated in Fig.~\ref{charect}(a,b), the even number of Co atoms form fully compensated spin sublattices connected through $C_{6z}$ screw-rotation and $M_z$ glide-mirror symmetries \cite{Dale} rather than inversion symmetry due to the distorted Nb-Se coordination environment [Fig.~S2], satisfying the established symmetry criteria for altermagnetism in CoNb$_4$Se$_8$ \cite{PhysRevX.12.031042,PhysRevX.12.040501}. This symmetry configuration satisfies the criteria for an altermagnetic state, in which compensated magnetic sublattices break time-reversal symmetry without generating a net macroscopic magnetization and produce momentum-dependent spin splitting even in the absence of SOC [Fig.~\ref{charect}(c)]. 
Consistent with this symmetry-governed electronic structure, CoNb$_4$Se$_8$ exhibits magnetic and transport properties characteristic of the previously reported altermagnetic state \cite{Regmi,Sakhya,Candelora,Dale,Devita}. Magnetization measurements under both zero-field-cooled and field-cooled conditions reveal an anomaly near 175 K [Fig.~\ref{charect}(d)], in agreement with the reported Néel temperature ($T_{\mathrm{N}}$) \cite{Regmi,Mandujano}, while the field response exhibits pronounced magnetic anisotropy between the $c$ axis and the $ab$ plane [Fig.~S3]. Microscopically, the collinear $A$-type antiferromagnetic order has been linked to the vertical arrangement of Co moments and their hybridization with Nb $4d_{z^2}$ orbitals \cite{Mandujano}. Consistently, the zero-field longitudinal resistivity ($\rho_\text{xx}$) displays a change in slope near $T_{\mathrm{N}}$ [Fig.~\ref{charect}(e)], indicating coupling between magnetic and electronic degrees of freedom, while the linear Hall ($\rho_\text{xy}$) response undergoes a sign reversal across $T_{\mathrm{N}}$ [Fig.~\ref{charect}(f)], further supporting the onset of magnetic ordering. 
Together, these results 
provide a suitable platform for investigating spin-lattice coupling, where phonon dynamics can reflect the underlying symmetry-governed magnetic interactions.

\subsection{Phonon dispersion}
\begin{figure}
\centering
\includegraphics[width=1\columnwidth]{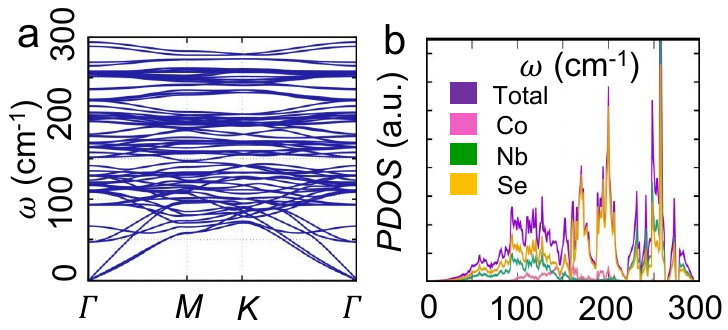}
\centering
\caption{
(a) Phonon dispersion relations and the corresponding (b) atom-projected phonon density of states (PDOS) of CoNb$_4$Se$_8$.} 
\vspace{-0.45cm}
\label{Raman_calculation}
\end{figure}

Figure~\ref{Raman_calculation} presents the calculated phonon dispersion and atom-projected phonon density of states (PDOS) for the magnetically ordered phase. The absence of imaginary modes across the Brillouin zone [Fig.~\ref{Raman_calculation}(a)] confirms dynamical stability \cite{Wei2025}, indicating that the magnetic transition does not induce structural symmetry breaking. 
The PDOS [Fig.~\ref{Raman_calculation}(b)] reveals a distinct separation of vibrational energy scales, where the low-frequency modes ($\lesssim$ $\sim$ 150 cm$^{-1}$) receive contributions from all three constituent atoms, while the higher-frequency modes ($\gtrsim$ $\sim$ 150 cm$^{-1}$) are predominantly dominated by the chalcogen atoms. 
Such separation can facilitate mode-dependent modulation of inequivalent exchange pathways through spin-phonon coupling, although it does not by itself determine the nature of the magnetic ground state. Rather, it provides a framework in which specific phonon modes couple selectively to distinct magnetic exchange channels, giving rise to symmetry-dependent renormalization of phonon energies and linewidths. 


\subsection{Selection rules for Raman processes}
Having established the dynamical stability [Fig.~\ref{Raman_calculation}(a)] of the magnetically ordered phase, we next identify the phonon modes relevant for Raman scattering using a symmetry-resolved lattice-dynamical analysis. 
To define the symmetry framework underlying spin-lattice interplay, we analyze the vibrational modes within the $D_{6h}$ ($6/mmm$) point group. Here, the primitive unit cell of CoNb$_4$Se$_8$ ($P6_3/mmc$, No.~194) contains 26 atoms (2 Co, 8 Nb, and 16 Se), corresponding to $Z=2$ formula units per cell. Group-theoretical decomposition yields a total of 78 normal modes (at the $\Gamma$ point),

\begin{equation}
\begin{aligned}
\Gamma_{\mathrm{vib}} =&\ 6A_{2u} \oplus 8E_{1u} \oplus 7E_{2g}
\oplus 6B_{1g} \oplus 2A_{2g} \oplus 6E_{2u} \\
&\oplus 5B_{2u} \oplus 4A_{1g}
\oplus 5E_{1g} \oplus B_{2g} \oplus A_{1u}.
\end{aligned}
\end{equation}

Within this symmetry, Raman activity is restricted to the $A_{1g}$, $E_{1g}$, and $E_{2g}$ representations, while infrared-active modes transform as $A_{2u}$ and $E_{1u}$; all remaining modes are optically silent. The coexistence of non-degenerate $A$- and $B$-type modes and doubly degenerate $E$ modes allows tracking of phonon renormalization for modes belonging to different symmetry representations across the magnetic transition. 
The $A$ and $E$ symmetry modes represent out-of-plane
and in-plane vibrations, respectively. In total, 28 Raman-active, 22 infrared-active, and 28 silent phonon modes (enlisted in Table~S3) are expected from symmetry analysis, providing a framework for mode-resolved spectroscopic investigation of spin–lattice coupling in this compensated magnetic system. 
Guided by the symmetry-derived selection rules and first-principles identification of Raman-active phonons, we now experimentally probe the lattice dynamics using Raman spectroscopy.

\begin{table*}
\caption{Calculated and experimental zone-center phonon frequencies at the $\Gamma$ point for CoNb$_4$Se$_8$. Experimental frequencies are obtained from Lorentzian fits to unpolarized Raman spectra measured with 532 nm laser excitation, while theoretical values are calculated with (w/) and without (w/o) SOC. The deviation from the expected anharmonic phonon frequency behavior is quantified by $\Delta \omega$ obtained from the temperature-dependent analysis presented in Fig.~\ref{2D_plot}.}
\centering
\setlength{\tabcolsep}{5pt}
\begin{tabular}{cccccc}
\hline\hline
Mode No. &
Freq. (cm$^{-1}$) &
Freq. (cm$^{-1}$) &
Freq. (cm$^{-1}$) &
Freq. (cm$^{-1}$) &
$\Delta \omega$ (cm$^{-1}$) \\
 & w/o SOC & w/ SOC & Exp. (300 K) & Exp. (80 K) & \\
\hline

P$_1$ & 95.8  & 93.21  & 99.9 $\pm$ 0.03 & 100.2 $\pm$ 0.06 & $\approx$ 0.01 \\
P$_2$ & 128.7 & 128.50 & 127.5 $\pm$ 0.15 & 131.2 $\pm$ 0.13 & $\approx$ 0.71 \\
P$_3$ & 136.4 & 136.05 & 133.3 $\pm$ 0.03 & 139.1 $\pm$ 0.02 & $\approx$ 0.87 \\
P$_4$ & 213.2 & 223.92 & 216.4 $\pm$ 2.31 & 208.6 $\pm$ 1.14 & -- \\
P$_5$ & 222.0 & 237.53 & 226.5 $\pm$ 0.05 & 228.4 $\pm$ 0.05 & $\approx$ 1.15 \\
P$_6$ & 240.4 & 252.29 & --     & -- & -- \\
P$^*$ & 243.7 & 256.60 & --     & -- & -- \\
P$_7$ & 257.0 & 257.48 & 255.8 $\pm$ 0.05 & 259.4 $\pm$ 0.04 & $\approx$ 0.01 \\
P$_8$ & 260.4 & 260.58 & 268.8 $\pm$ 0.13 & 273.5 $\pm$ 0.10 & $\approx$ 1.04 \\
\hline\hline
\end{tabular}
\label{phonon_cal}
\end{table*}


\subsection{Unpolarized Raman response} 
\begin{figure*}
\centering
\includegraphics[width=1.7\columnwidth]{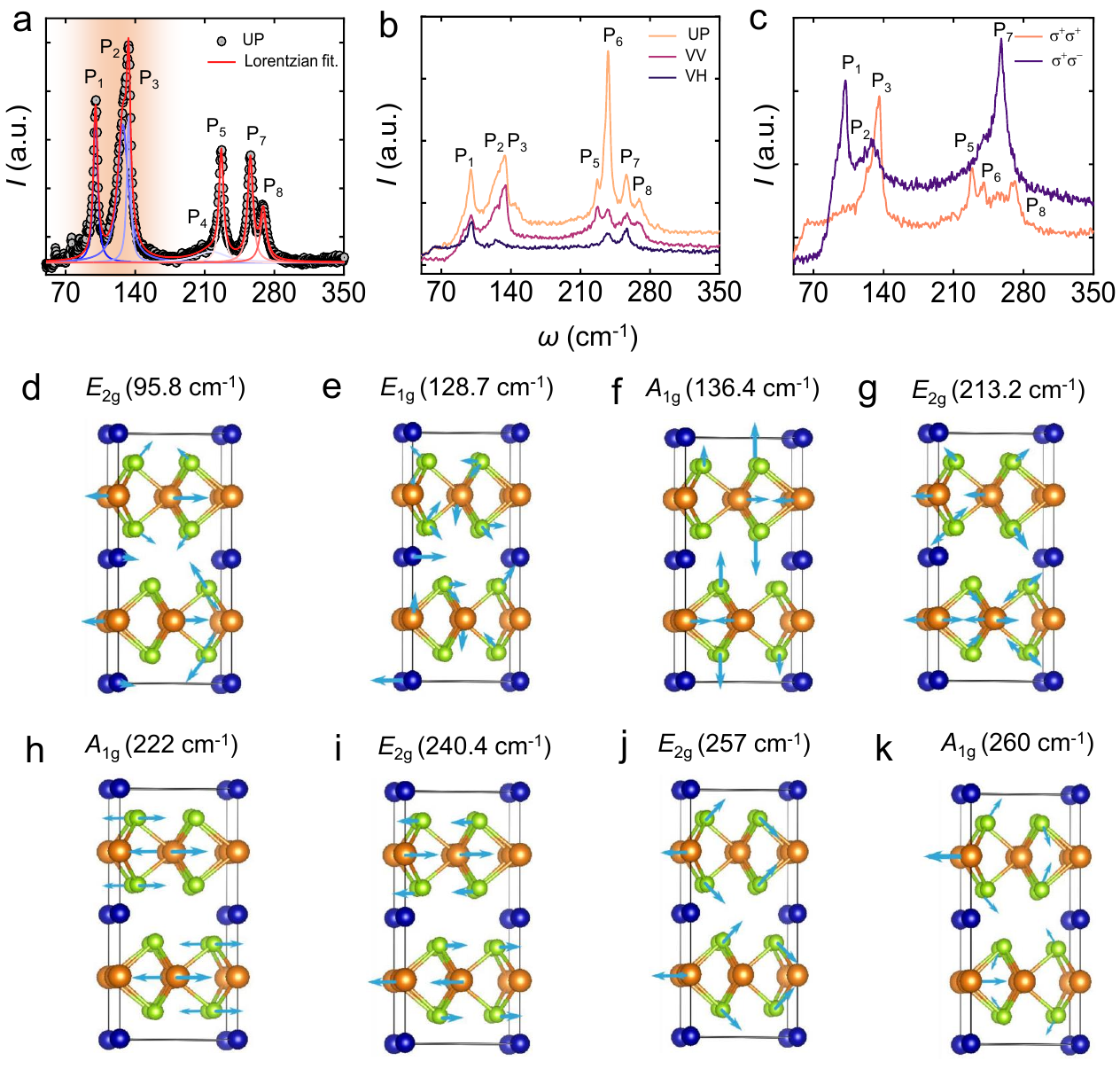}
\caption{
(a) 
Raman spectra of CoNb$_4$Se$_8$ acquired at room temperature using 532~nm excitation, resolving seven phonon modes (P$_1$ - P$_8$). The low-energy modes P$_1$–P$_3$, highlighted in orange, are attributed to superlattice-derived features. (b) Polarization-resolved Raman spectra measured at room temperature with 488~nm excitation in 
parallel (VV), and crossed (VH) linear configurations, along with (c) co- ($\sigma^+\sigma^+$) and cross ($\sigma^+\sigma^-$) -circular polarization configurations. (d-k) Calculated vibrational eigenmodes corresponding to the Raman-active phonon modes. Arrows indicate the direction and relative amplitude of atomic displacements. The color scheme denotes atomic species.}
\vspace{-0.45cm}
\label{Raman_T}
\end{figure*}

Figure~\ref{Raman_T}(a) shows the room-temperature Raman spectrum of CoNb$_4$Se$_8$ measured in a backscattering geometry, with 532 nm laser excitation. Multiple symmetry-allowed Raman-active modes are clearly resolved, spanning both low- and high-frequency regions. All observed phonon modes, labeled as P$_1$ - P$_8$, were quantitatively analyzed using multi-peak Lorentzian fitting. The experimental phonon assignments are further supported by first-principles lattice-dynamical calculations (see Table~\ref{phonon_cal}). 

We have P$_1$ - P$_8$ modes, which vibrate in the following manners: P$_1$ (95.8 cm$^{-1}$) [Fig.~\ref{Raman_T}(d)], P$_2$ (128.7 cm$^{-1}$) [Fig.~\ref{Raman_T}(e)], and P$_3$ (136.4 cm$^{-1}$) [Fig.~\ref{Raman_T}(f)] are characterized by highly anisotropic vibrations of the Se atoms. In these modes, the Co atoms exhibit in-plane shear vibrations. The Nb atoms undergo in-plane vibrations in the P$_1$ and P$_2$ modes, whereas P$_3$ involves out-of-plane Nb vibrations. Notably, the Nb atoms within a single layer vibrate antiparallel along the out-of-plane direction, while the corresponding interlayer Nb vibrations remain parallel and in phase. The P$_4$ (213.2 cm$^{-1}$) [Fig.~\ref{Raman_T}(g)] mode consists of out-of-plane vibrations of Co atoms in a breathing manner. All of the Se atoms of one layer vibrate in a diverging pattern, while the Se atoms of the other layer vibrate in a converging pattern. Specifically, P$_5$ at 222 cm$^{-1}$ [Fig.~\ref{Raman_T}(h)] corresponds to out-of-plane vibrations of both Nb and Se atoms across the layers, whereas the higher-frequency peak P$_7$ (257 cm$^{-1}$)[Fig.~\ref{Raman_T}(j)] 
characterized by coherent in-plane Nb motion accompanied by stretching-contraction dynamics of the Se framework. 
Moreover, The P$_8$ (260.0 cm$^{-1}$) [Fig.~\ref{Raman_T}(k)] mode has in-plane shear vibrations of Nb atoms, while the Se atoms have mixed diverging and converging vibrations. Co atoms have negligible vibrations for this mode. 

Although pristine NbSe$_2$ and CoNb$_4$Se$_8$ belong to the same space group, the latter exhibits a richer Raman spectrum with a larger number of Raman-active modes. This behavior originates from its larger primitive cell, lower site symmetry, and the presence of Co-Se vibrations \cite{Erodici}. By contrast, NbSe$_2$ is characterized by only two Raman-active phonon modes near 250 cm$^{-1}$ and 225 cm$^{-1}$ \cite{Xi}.

While unpolarized Raman spectra provide an overview of the phonon landscape, polarization-resolved measurements are essential for disentangling the symmetry and selection rules governing the observed modes.

\subsection{Polarization-resolved Raman spectra} -- 
Figure~\ref{Raman_T}(b) presents the Raman spectra of CoNb$_4$Se$_8$ measured at room temperature using 488 nm excitation under linear polarization (LP), including parallel (VV) and crossed (VH) scattering geometries, along with the unpolarized (UP) response for comparison. The pronounced polarization dependence of the phonon intensities reflects well-defined Raman selection rules, indicative of 
anisotropic lattice dynamics consistent with the $P6_3/mmc$ ($D_{6h}$) symmetry \cite{Bilbao}. In the VV configuration (incident light and scattered light have the same polarization direction, $0^\circ$), both $A_\text{1g}$ and $E_\text{2g}$ modes are observed due to finite diagonal Raman tensor elements, 
whereas the VH geometry (incident light and scattered light are perpendicular, $90^\circ$) selectively probes off-diagonal components, 
resulting in the dominance of $E_\text{2g}$ modes, in full agreement with symmetry-imposed selection rules \cite{Bilbao}. Accordingly, modes P$_1$, P$_2$, P$_6$, and P$_7$ are clearly observed in both VV and VH channels, confirming their $E_\text{2g}$ character, while modes P$_3$, P$_5$, and P$_8$ are predominantly confined to the VV configuration, consistent with their assignment to $A_\text{1g}$ symmetry. The detailed phonon frequencies and their polarization dependence are summarized in Table~\ref{phonon_symmetry}, showing excellent agreement between calculated and experimentally observed symmetries. As the relative polarization angle is varied from $0^\circ$ to $90^\circ$, the intensity of the $A_\text{1g}$ modes is progressively suppressed, whereas the $E_\text{2g}$ modes remain robust and, in some cases, become more prominent. In particular, the mode near $\sim 236~\mathrm{cm^{-1}}$ (P$_6$) [Fig.~\ref{Raman_T}(i)] exhibits strong intensity in both VV and VH geometries, consistent with its $E_\text{2g}$ symmetry, and corresponds to predominantly in-plane vibrations of Nb and Se atoms with the two layers vibrating in opposite directions, while the Co atoms remain nearly stationary. 

 Helicity-resolved Raman spectra [Fig.~\ref{Raman_T}(c)] exhibit a polarization dependence consistent with the linearly polarized measurements. Notably, the doubly degenerate $E_\text{2g}$ modes remain active in both the helicity-preserving co-circular $(\sigma^{+}\sigma^{+})$ and helicity-flipping cross-circular $(\sigma^{+}\sigma^{-})$ configurations, consistent with their distinct Raman tensor symmetry and in-plane vibrational character. In contrast, the nondegenerate modes are largely suppressed in the cross-circular geometry and predominantly remain active only in the co-circular channel. These helicity-dependent selection rules reflect the symmetry-selective nature of circularly polarized Raman scattering and provide additional information on the symmetry character of the vibrational modes in CoNb$_4$Se$_8$. 
 \begin{figure}
\centering
\includegraphics[width=1\columnwidth]{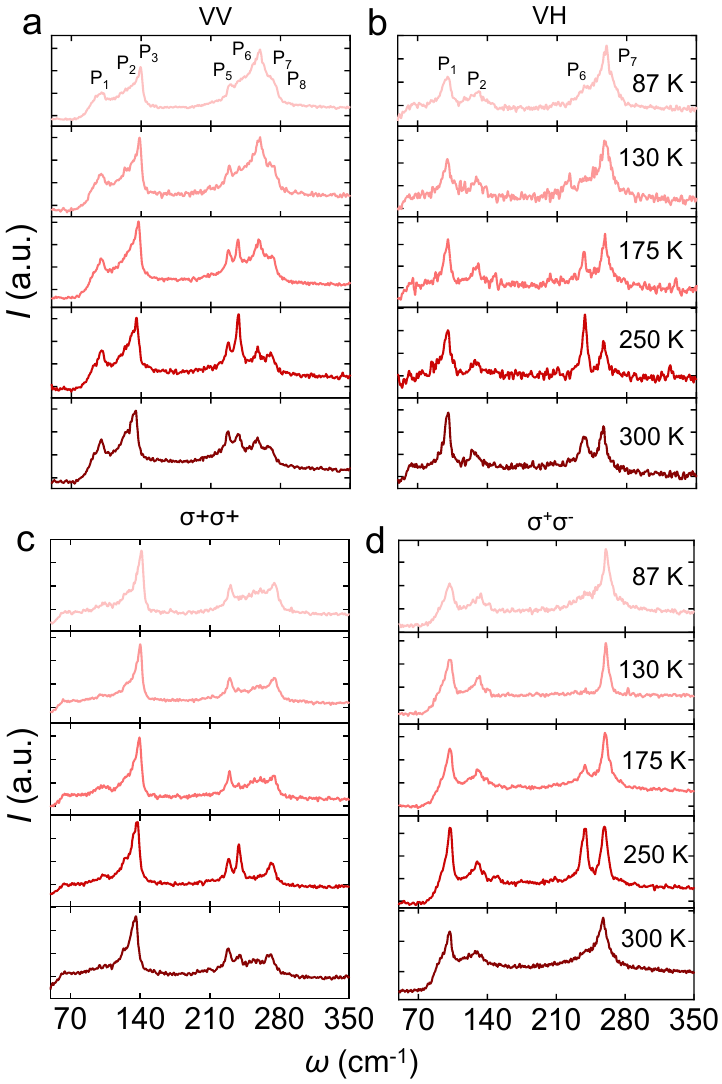}
\caption{Temperature-dependent Raman spectra of CoNb$_4$Se$_8$ from 87~K to 300~K showing the evolution of phonon modes P$_1$ - P$_8$: (a) VV, and (b) VH linear polarization geometries, and  (c) $\sigma^+\sigma^+$ and (d) $\sigma^+\sigma^-$ circular configurations.} 
\vspace{-0.45cm}
\label{LP_CP}
\end{figure}
\begin{figure*}
\centering
\includegraphics[width=1.69\columnwidth]{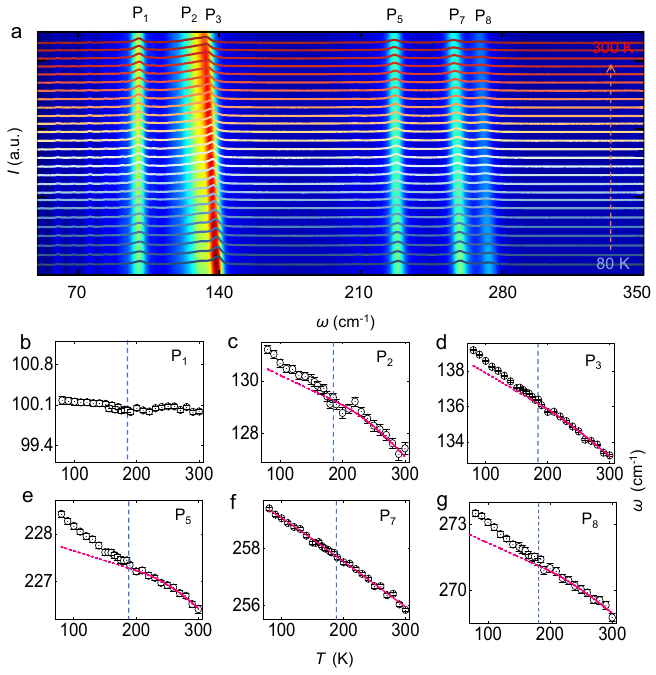}
\caption{
(a) False-color map of Raman intensity in the 50–350~cm$^{-1}$ range as a function of temperature (80–300~K) for phonon modes P$_1$ - P$_8$. The orange dashed arrow indicates increasing temperature. The P$_4$ mode is weak and not clearly resolved due to its comparatively low intensity.
(b-g) Temperature dependence of phonon frequencies ($\omega$) from 80 to 300~K. The blue dashed line marks the Néel temperature ($T_{\mathrm{N}}$), while the pink solid lines represent fits based on anharmonic phonon--phonon scattering theory~\cite{Balkanski}, parameterized in the high-temperature regime ($T > T_{\mathrm{N}}$).} 
\vspace{-0.45cm}
\label{2D_plot}
\end{figure*}
\subsection{Unified structural symmetry}
Building on the polarization-resolved Raman response at room temperature, we examine its temperature evolution across the magnetic transition. The set of phonon modes remains unchanged across $T_{\mathrm{N}}$ in both linear (VV, VH) [Fig.~\ref{LP_CP}(a,b)] and circular ($\sigma^+\sigma^+$, $\sigma^+\sigma^-$) [Fig.~\ref{LP_CP}(c,d)] configurations, with no emergence or disappearance of peaks, indicating that the selection rules are preserved and no structural symmetry change occurs within experimental resolution. Subtle variations in intensity and linewidth are observed for selected modes near $T_{\mathrm{N}}$, reflecting the influence of magnetic ordering on lattice vibrations. The persistence of helicity-dependent differences in circular polarization further indicates the sensitivity of the phonon response to the magnetic state. 
Overall, these results indicate a possible coupling between lattice vibrations and the underlying spin degrees of freedom, without evidence for a change in crystallographic symmetry.

\subsection{Temperature-dependent 
Raman spectra} Temperature-dependent Raman measurements were performed over the range 78–300~K [Fig.~\ref{2D_plot}(a)] to examine the evolution of lattice dynamics across the magnetic transition. The temperature-dependent results discussed below are derived from unpolarized Raman spectra, which provide a comprehensive reference for phonon-mode identification by simultaneously probing multiple symmetry channels without imposing polarization selection rules. 
The Raman-active modes (P$_1$ - P$_8$) are tracked as a function of temperature, exhibiting mode-dependent variations in frequency and linewidth. While the overall behavior is consistent with anharmonic lattice dynamics, small deviations are observed near $T_{\mathrm{N}}$, suggesting a possible influence of magnetic ordering on the lattice vibrations.

The low-energy $E_\mathrm{2g}$ mode P$_1$ exhibits an almost temperature-independent frequency [Fig.~\ref{2D_plot}(b)], with only a slight redshift upon warming that is well described by anharmonic phonon behavior. A barely discernible softening near $T_\mathrm{N}$ suggests a weak magnetic renormalization of the phonon self-energy. 
In contrast, the intermediate-frequency mode P$_2$ shows a pronounced nonlinear redshift [Fig.~\ref{2D_plot}(c)] that deviates from the standard three-phonon anharmonic model \cite{Balkanski}, suggesting an additional contribution to the phonon self-energy.

The $A_\text{1g}$ phonon modes, particularly P$_3$, P$_5$, and P$_8$ [Figs.~\ref{2D_plot}(d), \ref{2D_plot}(e), and \ref{2D_plot}(g)], exhibit substantially stronger frequency renormalization and larger deviations from the anharmonic trend than the $E_\mathrm{2g}$ modes. In contrast, the in-plane $E_\mathrm{2g}$ mode P$_7$ follows a nearly conventional anharmonic temperature evolution [Fig.~\ref{2D_plot}(f)]. Among all modes, P$_5$ exhibits the largest anharmonic deviation ($\Delta\omega$) [Table~\ref{phonon_cal}], highlighting its enhanced sensitivity to the evolving magnetic environment. The contrasting responses of the $A_\text{1g}$ and $E_\mathrm{2g}$ phonons therefore point to a symmetry-selective coupling mechanism that cannot be fully captured within a purely phonon-phonon or conventional exchange-driven framework.


Overall, the temperature evolution is characterized by a gradual softening and linewidth broadening [Fig.~S6(a,b)] of phonon modes, arising from lattice anharmonicity and thermal expansion of the lattice~\cite{Zhu2021}. Within this overall trend, certain modes, particularly P$_4$, exhibit comparatively stronger renormalization and display an anomalous blue shift with increasing temperature [Fig.~S6(c)]. Such behavior may reflect differences in volumetric expansion between in-plane and out-of-plane intercalant-related vibrational modes \cite{Zhu2021,Steurer}.

A closer examination of the temperature evolution of individual phonon modes provides insight into the underlying mechanisms governing their renormalization. The absence of a universal scaling of phonon shifts with temperature across all modes suggests that the observed renormalization cannot be described solely by anharmonic effects. Instead, the mode-selective behavior indicates symmetry-dependent interactions that modulate the phonon self-energy. While electron-phonon coupling cannot be completely excluded, the lack of Fano asymmetry in the Raman line shapes [see Section S7(B) for details] and the $T^2$ dependence of resistivity [Fig.~S4(a)] suggest that it is not the dominant mechanism. These observations are therefore consistent with a significant contribution from spin-phonon coupling, governed by symmetry-selective interactions in the system.

\subsection{Spin-phonon coupling beyond long-range magnetic order} 
\begin{figure}[t]
\centering
\includegraphics[width=1\columnwidth]{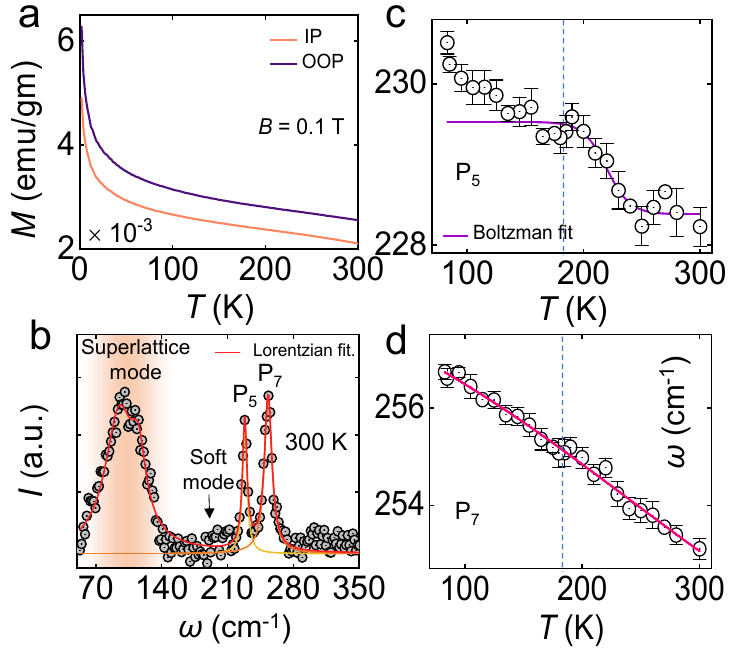}
\caption{
(a) Temperature dependence of magnetization measured at $B = 0.1$~T for in-plane (IP) and out-of-plane (OOP) configurations.
(b) Unpolarized Raman spectrum at room temperature, showing three distinct phonon features (superlattice peak, P$_5$, and P$_7$).
(c,d) Temperature evolution of the phonon peak positions for modes P$_5$ and P$_7$, respectively. The violet solid line in (c) represents a fit using a Boltzmann sigmoidal function.} 
\vspace{-0.45cm}
\label{charect_co}
\end{figure}

To further elucidate the origin of the observed phonon anomalies, we conducted a comparative study of a Co-deficient analogue synthesized using the same growth protocol as CoNb$_4$Se$_8$, but with a modified stoichiometric ratio. Such a comparison enables us to distinguish lattice responses associated with coherent long-range magnetic order from those linked to local symmetry-governed magnetic correlations that may persist even when long-range order is weakened. For this purpose, Co$_{0.57}$Nb$_4$Se$_8$ was employed as a representative reference system. This compound preserves the same crystallographic framework [Fig.~S7(a,b)] but exhibits only short-range antiferromagnetic correlations [Fig.~S7(c)], without clear signatures of long-range magnetic order \cite{Mandujano} associated with an altermagnetic state [Fig.~\ref{charect_co}(a)]. Room-temperature Raman spectra reveal the P$_5$ and P$_7$ phonon modes together with a broad superlattice-related feature [Fig.~\ref{charect_co}(b), Table~S4], while the corresponding linearly and circularly polarized spectra are presented in the Supplementary Information [Fig.~S8(a,b)]. 
Both phonon modes exhibit pronounced temperature-dependent evolution, with distinct anomalies emerging around 180 K, close to the Néel temperature of CoNb$_4$Se$_8$ (Fig.~S9). These anomalies are manifested through changes in slope and enhanced phonon frequency renormalization across the transition region. Notably, the P$_5$ mode displays a pronounced deviation across this temperature scale [Fig.~\ref{charect_co}(c)], closely resembling the behavior observed in the pristine compound [Fig.~\ref{Raman_T}(h)], thereby indicating a similar sensitivity to the underlying spin correlations. In contrast, the P$_7$ mode evolves smoothly over the entire temperature range without exhibiting a distinct anomaly near $T_\mathrm{N}$ [Fig.~\ref{charect_co}(d) and Fig.~\ref{Raman_T}(j)], consistent with its comparatively weaker coupling to the magnetic sublattices, as also observed in the stoichiometric system.

\subsection{Discussion} The temperature- and polarization-dependent Raman measurements reveal a strongly mode-dependent phonon response. Specifically, P$_5$ exhibit noticeable deviations from the expected anharmonic behavior near $T_{\mathrm{N}}$ (Table~\ref{phonon_cal}), P$_6$ displays a pronounced spectral-weight redistribution across the transition [Fig.~\ref{LP_CP}], and P$_7$ remains largely unaffected [Fig.~\ref{2D_plot}(f) and Fig.~\ref{charect_co}(d)]. This suggests that the phonon
self-energy may acquire symmetry-dependent contributions,
influenced not only by interatomic distances but
also by the magnetic configuration. Such observations
indicate that a purely scalar description of spin-phonon
coupling may not fully capture the observed phonon behavior.

In this context, spin-orbit coupling, although not the primary origin of spin polarization in these systems \cite{PhysRevX.12.040501}, provides a possible microscopic pathway linking lattice, orbital, and spin degrees of freedom \cite{Schilberth2026}. Through coupling between orbital character and spin polarization, SOC can allow lattice distortions to influence the spin-dependent electronic structure in a symmetry-dependent manner, thereby contributing to phonon renormalization. First-principles calculations support this picture: While calculations without SOC establish the baseline phonon spectrum, the inclusion of SOC induces selective phonon renormalization for several branches (P$_4$, P$_5$, P$_6$, and P$^*$; Table~\ref{phonon_cal}), consistent with the experimentally observed symmetry-selective phonon response. Together with the Raman measurements, these results suggest that SOC influences different phonon symmetries unequally, providing a plausible microscopic origin for the observed symmetry-dependent spin-lattice coupling. There have been reports highlighting this indirect way of realizing SOC for various materials \cite{paul2015spin,kim2020spin}. There are some rigorous theoretical approaches that can calculate the SPC coefficients \cite{lunghi2017intra, albino2021temperature, nabi2023accurate}, however, the case studies relevant to the altermagnets are not discussed systematically to the best of our knowledge. Although Nb and Se possess only moderate atomic numbers, their 4$d$ and 4$p$ electronic states provide a finite SOC strength \cite{Winkler2003}, which may become relevant in a symmetry-sensitive magnetic environment.

No additional phonon modes emerge or disappear across $T_\mathrm{N}$ [Fig.~\ref{LP_CP} and Fig.~\ref{2D_plot}(a)], consistent with the absence of a structural phase transition \cite{Regmi}. The observed anomalies therefore arise not from crystallographic symmetry breaking but from mode-dependent phonon renormalization, manifested through frequency shifts, linewidth changes, and spectral-weight redistribution. This behavior underscores the strong coupling between lattice dynamics and the evolving magnetic state while preserving the underlying crystal symmetry.

The persistence of comparable phonon characteristics even in the absence of well-defined long-range magnetic order [Fig.~\ref{charect_co}] further suggests that the coupling is not determined solely by coherent magnetic ordering. Instead, short-range spin correlations or local symmetry-constrained magnetic textures may also contribute to the observed renormalization, similar to behavior reported in frustrated magnetic systems where SPC survives above the Néel temperature \cite{Huang}. Within this framework, SOC act as an enabling interaction connecting lattice vibrations with the spin sector under symmetry constraints, rather than serving as the primary origin of magnetism itself \cite{LiU2026}.

A useful comparison can be made with Co$_2$Mo$_3$O$_8$, a predicted AM \cite{Schilberth2026}, where phonon renormalization across the antiferromagnetic transition has been associated with changes in Raman selection rules linked to relativistic magnetic symmetry. In contrast, the present system does not exhibit the appearance or suppression of symmetry-forbidden modes across $T_{\mathrm{N}}$ within experimental resolution, indicating that the Raman selection rules remain preserved. Instead, the phonon response is reflected primarily through mode-dependent renormalization and helicity-dependent intensity variations with temperature, demonstrating that lattice dynamics can remain sensitive to the magnetic environment without requiring a detectable change in crystal symmetry. Within this picture, temperature primarily influences short-range spin correlations and the underlying electronic environment, consistent with coupling effects extending beyond $T_{\mathrm{N}}$. The observed mode selectivity together with its polarization dependence further suggests that SOC-influenced interactions may contribute to the phonon response beyond a purely conventional exchange-driven description.

\begin{table*}
\caption{Experimental Raman-active phonon modes of CoNb$_4$Se$_8$ at 300 K and 87 K using 488 nm laser excitation, together with their experimentally observed (Obs.) and theoretically calculated (Cal.) symmetry assignments, obtained from linear-polarization (LP) and circular-polarization (CP) Raman scattering configurations.}
\centering
\setlength{\tabcolsep}{0.08pt}
\begin{tabular}{c cccc cccc cc}
\hline\hline
& \multicolumn{4}{c}{Linear Polarization (LP)} 
& \multicolumn{4}{c}{Circular Polarization (CP)} 
& \multicolumn{2}{c}{Symmetry} \\
\cline{2-5} \cline{6-9} \cline{10-11}
Mode 
& \multicolumn{2}{c}{VH} 
& \multicolumn{2}{c}{VV} 
& \multicolumn{2}{c}{$\sigma^+\sigma^-$} 
& \multicolumn{2}{c}{$\sigma^+\sigma^+$} 
& (Cal.) & (Obs.) \\
& 300 K & 87 K & 300 K & 87 K 
& 300 K & 87 K & 300 K & 87 K & & \\
\hline

P$_1$ & 99.6 $\pm$ 0.07  & 98.9 $\pm$ 0.17 & 98.7 $\pm$ 0.19 & 98.9 $\pm$ 0.19 & 99.8 $\pm$ 0.07 & 100.9 $\pm$ 0.11 & 102.5 $\pm$ 1.02 & 104.7 $\pm$ 0.71 & $E_{2g}$ & $E_{2g}$ \\
P$_2$ & 126.7 $\pm$ 0.35 & 129.5 $\pm$ 0.39 & 125.0 $\pm$ 0.39 & 127.1 $\pm$ 0.30 & 127.4 $\pm$ 0.23 & 131.0 $\pm$ 0.28 & 126.0 $\pm$ 0.48 & 130.3 $\pm$ 0.42 & $E_{2g}$ & $E_{2g}$ \\
P$_3$ & -- & -- & 133.6 $\pm$ 0.08 & 138.8 $\pm$ 0.06 & -- & -- & 134.8 $\pm$ 0.06 & 140.6 $\pm$ 0.05 & $A_{1g}$ & $A_{1g}$ \\
P$_4$ & -- & -- & -- & -- & -- & -- & -- & -- & $E_{2g}$ & -- \\
P$_5$ & -- & -- & 227.3 $\pm$ 0.37 & 229.0 $\pm$ 0.22 & -- & -- & 228.3 $\pm$ 0.11 & 230.9 $\pm$ 0.20 & $A_{1g}$ & $A_{1g}$ \\
P$_6$ & 237.0 $\pm$ 0.22 & 240.1 $\pm$ 0.80 & 238.1 $\pm$ 0.18 & 241.9 $\pm$ 0.48 & 240.2 $\pm$ 0.69 & 250.4 $\pm$ 0.73 & 240.1 $\pm$ 0.90 & -- & $E_{2g}$ & $E_{2g}$ \\
P$_7$ & 256.1 $\pm$ 0.12 & 259.1 $\pm$ 0.13 & 256.0 $\pm$ 0.29 & 258.6 $\pm$ 0.12 & 257.5 $\pm$ 0.08 & 261.1 $\pm$ 0.07 & 256.5 $\pm$ 0.12 & 257.4 $\pm$ 0.47 & $E_{2g}$ & $E_{2g}$ \\
P$_8$ & -- & -- & 269.3 $\pm$ 0.28 & 271.9 $\pm$ 0.20 & -- & -- & 270.3 $\pm$ 0.18 & 274.4 $\pm$ 0.15 & $A_{1g}$ & $A_{1g}$ \\

\hline\hline
\end{tabular}
\label{phonon_symmetry}
\end{table*}

\section{Outlook \& Conclusions}
To investigate the origin of spin-phonon coupling in symmetry-governed magnetic systems, where momentum-space spin polarization arises from crystal symmetry rather than net magnetization, we perform a comparative study of the altermagnetic metal CoNb$_4$Se$_8$ and a structurally analogous Co-deficient compound. Raman spectroscopy reveals pronounced symmetry-selective phonon renormalization across the magnetic transition in CoNb$_4$Se$_8$, while similar anomalies persist even when long-range magnetic order is significantly suppressed. This indicates that the observed spin-lattice response cannot be accounted for solely by conventional exchange-striction associated with coherent magnetic ordering. First-principles calculations show that spin-orbit coupling provides a symmetry-dependent channel that links lattice vibrations to the spin-polarized electronic structure. The phonon modes most affected in the calculations are consistent with those exhibiting the strongest experimental anomalies, supporting a direct correspondence between theory and experiment. Overall, these results suggest an additional contribution to spin-phonon coupling in symmetry-governed magnetic materials and highlight the sensitivity of lattice dynamics to symmetry-derived spin-polarized states. More broadly, this work provides a basis for understanding spin-lattice interactions in these systems and may guide future efforts toward controlling spin-dependent properties through crystal symmetry and spin-orbit interactions.
\section{Acknowledgment}
SKP acknowledges the Saha Institute of Nuclear Physics (SINP) for financial support through a Postdoctoral Fellowship and for providing access to the experimental facilities used in this work. DD acknowledges the University Grants Commission (UGC), India, for financial support through a fellowship. SN acknowledges financial support from the Anusandhan National Research Foundation (ANRF), Government of India, through the Prime Minister’s Early Career Research Grant (ANRF/ECRG/2024/005947/PMS). SDG acknowledges funding support from the Department of Atomic Energy (DAE) and the Anusandhan National Research Foundation (ANRF) through the Prime Minister’s Early Career Research Grant (ANRF/ECRG/2024/004649/PMS). ANP acknowledges funding from DST Nano Mission: Grant no. DST/NM/TUE/QM-10/2019.
\section{Author contributions}
SKP and ANP conceived the project. SKP, SM, DM and AB synthesized the single-crystals and carried out structural, magnetic and magnetotransport experiments with inputs from SDG. SKP and DD carried out unpolarized Raman spectroscopy experiments with inputs from ANP.  SM and SKP carried out polarized Raman spectroscopy experiments with inputs from AS. SP and SN carried out first-principles calculations. SKP wrote the manuscript with feedback from AS, SN, ANP and SDG. All authors reviewed and approved the final version of the manuscript.

\section{Conflict of Interest}
The authors declare no conflict of interest.

\section{Data Availability Statement}
The data that support the findings of this study are available from the corresponding author upon reasonable request.

\section{Keywords} Altermagnet; Spin-phonon coupling; Spin-orbit coupling; Raman spectroscopy; Lattice dynamics; Quantum materials.

\section{Methods}

\textbf{Crystal growth} — Single crystals of CoNb$_4$Se$_8$ were first synthesized using a two-step solid-state reaction followed by chemical vapor transport (CVT), following the procedure reported in Ref.~\cite{Regmi}. Subsequently, Co-deficient composition was grown using the same synthesis protocol, but with 
a modified stoichiometric ratio.

\textbf{Structural and compositional characterization} — Crystal orientation and phase purity were characterized by room-temperature powder X-ray diffraction (PXRD) using a Rigaku SmartLab diffractometer with Cu K$\alpha$ radiation. Rietveld refinement was performed using the MAUD software package \cite{Maudsoftware}, and structural visualization was carried out using VESTA \cite{Momma:db5098}. Elemental composition was determined by energy-dispersive X-ray spectroscopy (EDX) using a QUANTA field-emission-gun 250 scanning electron microscope on multiple crystals from the same growth batch.

\textbf{Magnetic and transport measurements} — Magnetic, electronic, and magnetotransport measurements were performed using a Quantum Design physical property measurement system (PPMS) in the temperature range 5–300 K. Temperature-dependent magnetization was measured under zero-field-cooled (ZFC) and field-cooled (FC) conditions with magnetic fields applied both parallel and perpendicular to the sample surface. Electrical transport measurements were carried out in a standard four-probe geometry with a ${dc}$ current applied within the $ab$ plane. Longitudinal ($\rho_{\mathrm{xx}}$) and transverse ($\rho_{\mathrm{xy}}$) resistivities were measured as functions of temperature and magnetic field. 

\textbf{Raman spectroscopy} — Unpolarized Raman measurements were performed in back-scattering geometry using a HORIBA LABRAM HR spectrometer with 532 and 633 nm excitation sources. Polarization-resolved linear and circular Raman measurements were carried out using a Jobin Yvon LabRAM HR micro-Raman system equipped with polarization optics, including linear polarizers, analyzers, and a $\lambda/4$ plate, using a 488 nm Ar laser excitation source. For all measurements, spectra were recorded using an 1800 gr/mm grating, providing a spectral resolution of approximately 0.5 cm$^{-1}$. The experimental Raman setup schematic is shown in Fig.~S5. Temperature-dependent measurements were conducted using a Linkam THMS600 stage under vacuum conditions. The laser power was kept below 0.2 mW to avoid sample heating. Prior to measurements, the optical path was calibrated using a single-crystal Si standard. Thick exfoliated single crystals placed on Si(100) substrates were used to ensure high Raman signal quality.

\textbf{First-principles calculations} — Electronic structure and lattice-dynamics calculations were performed within density functional theory using the {\footnotesize QUANTUM ESPRESSO} package \cite{giannozzi2009quantum,giannozzi2017advanced,giannozzi2020quantum} with norm-conserving pseudopotentials \cite{troullier1991efficient}. A plane-wave cutoff energy of 70 Ry and Methfessel–Paxton smearing of 0.01 Ry were employed. Phonon dispersions were obtained by Fourier interpolation of dynamical matrices calculated on $6\times6\times6$ \textbf{k}- and $3\times3\times3$ \textbf{q}-point meshes.
\bibliographystyle{apsrev4-2}
\bibliography{maintext}{}	
\end{document}